# Investigation of Superconducting Gap Structure in TbFeAsO$_{0.9}$F$_{0.1}$ using Point Contact Andreev Reflection

K A Yates[1,3], K. Morrison[1], Jennifer A. Rodgers[2], George B. S. Penny[2], Jan-Willem G. Bos[2], J. Paul Attfield[2], and L F Cohen[1]

[1] The Blackett Laboratory, Physics Department, Imperial College London, SW7 2AZ, UK

[2] Centre for Science at Extreme Conditions and School of Chemistry, University of Edinburgh, King's Buildings, Mayfield Road, Edinburgh, EH9 3JZ.

Bulk samples of TbFeAsO$_{0.9}$F$_{0.1}$ ($T_c^{on}$ = 50K) were measured by point contact Andreev reflection spectroscopy. The spectra show unambiguous evidence for multiple gap–like features plus the presence of high bias shoulders. By measuring the spectra as a function of temperature with both gold and superconducting niobium tips, we establish that the gap-like features are associated with superconducting order parameter in this material. We discuss whether the well defined zero bias conductance peak that we observe infrequently is associated with a nodal superconducting order parameter.

[3] Corresponding author, Karen A Yates, e-mail: k.yates@imperial.ac.uk

**1. Introduction**
Since the discovery of superconductivity at 26K in fluorine doped LaFeAsO$_{1-x}$F$_x$[1] there has been considerable experimental and theoretical work devoted to fundamental questions related to the nature of the superconductivity in these materials. Many theoretical models suggest that the gap order is unconventional with nodes in the gap structure[2,3], multigap extended s-wave symmetry[4] or some combination thereof. Experimentally, the order parameter has been interpreted as both nodeless[5,6,7,8,9] or nodal[10,11,12,13,14] and either single gap[9,12,13] or multiply gapped[8,14,15]. There is little consensus yet partly due to consistency across the samples, partly owing to the role played by rare earth substitution and the differences concerning doping with F or oxygen vacancies. At this early stage it is important that as many results as possible are shown and discussed so that some consensus can be established. In this paper, we present point contact Andreev reflection (PCAR) spectroscopy data as a function of magnetic field and temperature on high pressure synthesised samples of TbFeAsO$_{0.9}$F$_{0.1}$ in order to directly probe these questions. We also interrogate the sample using a superconducting niobium tip in order to examine whether the resulting sub-gap structure can validate the spectroscopic features as gap structures and whether the dimensionality of the associated Fermi surface sheets can be established.

**2. Experimental Method and Data Analysis**
The bulk, polycrystalline TbFeAsO$_{0.9}$F$_{0.1}$ samples studied here were prepared by a high pressure synthesis route as described in ref [16]. The sample was measured resistively and found to have a superconducting onset temperature of $T_c^{on}$ = 50K and a transition width $\Delta T_c$ ~ 3K. Point contact Andreev reflection measurements were taken with a mechanically sharpened tip of either Au or Nb as described in detail elsewhere[17]. Measurements were taken at 4.2 K when both tip and sample were immersed in liquid helium and as a function of temperature and magnetic field.
We use a four point method to determine the conductance spectra by sweeping the voltage bias across the tip-sample junction and measuring the differential conductance of the ac ripple applied on top of this bias sweep. The contact size assuming one single junction contact, calculated using the Sharvin formula[18] was found to be in the 3nm - 8nm range. Although this is comparable to the reported coherence length for the oxypnictides[19], it is much less than the mean free path of the gold tip. Moreover, as the junction between tip and sample is made up of many tens of parallel channels[23], this means that the individual contacts are well inside the ballistic regime and consequently heating effects across the junction can be neglected.
For the analysis of the PCAR data we use the Blonder, Tinkham, Klapwijk (BTK) formulism that results in the extraction of three parameters, the superconducting energy gap $\Delta$, the broadening parameter $\Gamma$ (which incorporates thermal and non thermal broadening contributions) and the interface parameter Z which is taken to be a delta function potential at the interface[20]. We extend this model in various ways depending on the nature of the spectroscopic data. For multiple gaps we assume that the conductance spectrum is a weighted sum of contributions from two superconducting order parameters in order to extract $\Delta_1$ and $\Delta_2$ [17]. For data taken in magnetic field we incorporate the two channel model[21] that allows a proper description for the influence of vortices (for a simple s wave superconductor), and for spectra that show a clear zero bias conductance peak we incorporate the model by Kashiwaya-Tanaka[22] for a $d_{x2-y2}$ order parameter that results in an extra variable being introduced to the model, $\alpha$, the angle between the quasiparticle injection trajectory and the superconducting order parameter antinode. Due to the introduction of a fourth fitting parameter and the associated possibility of degenerate fits[23], we adapted a $\chi^2(\alpha)$ fitting procedure that we have described in detail elsewhere (with respect to fitting the spin polarisation P of ferromagnetic materials in reference 23). In the adapted $\chi^2(\alpha)$ procedure used here, the value of $\alpha$ is varied incrementally and a three parameter fit (for $\Delta$, Z and $\Gamma$) is performed in order to calculate $\chi^2$. The resulting plot of $\chi^2$ vs $\alpha$ shows a minimum in $\chi^2(\alpha)$ that indicates the best fit to the data. However, it should be noted that in polycrystalline samples, $\alpha$ is expected to be an average value taken across several randomly oriented grains[24].

## 3. Results

Zero Bias Conductance Peak

We start by showing in figure 1(a) a typical example of spectroscopic data that shows the zbcp feature and how it evolves with temperature. The zbcp decreases in height with increasing temperature. Figure 1(b) shows the fit of the 4.2K spectra to the Kashiwaya-Tanaka $d_{x^2-y^2}$ model with $\Delta = 8.5$meV, $Z = 0.4$, $\alpha = 0.21$rad and $\Gamma = 0.97$meV. To fit the higher temperature spectra to the same model, both the Z and $\alpha$ parameters have to be adjusted systematically, (Z by 20% and $\alpha$ by 10%) which is rather unphysical given the fact that the contact resistance at high bias was constant with increasing temperature, suggesting a stable contact and therefore fixed values of these parameters. We can project how we expect the spectra to evolve as a function of temperature if subjected only to increased thermal broadening using the fitted values for the 4.2K data. In order to compare directly to experiment we show the ratio of the height of the zbcp compared to that of the finite bias conductance peak, figure 2. Clearly the experimental variation of the zbcp is inconsistent with a simple d-wave scenario to explain the zbcp. Samuely et al,[11] have already noted that the temperature dependence of the zbcp peak is unusual within the context of a simple d-wave superconductor, although the observed temperature dependence in that paper differs significantly from that reported here. Furthermore, we observe no splitting of the zbcp with magnetic field (figure 1c) as may be expected for a d-wave superconductor[25], although this may be a consequence of the orientation of the contact with respect to the superconductor crystal axis.

In order to explore the hypothesis that a Josephson junction in series with the contact is responsible for the zbcp[24] we study the field evolution of a different set of spectra that also show the zbcp anomaly. The results are shown in figure 1(c). A Josephson junction would be expected to show a rapid decrease in zbcp height with increasing magnetic field which is not observed. The spectra are therefore inconsistent with either a d-wave order parameter or a simple, single s-wave order parameter with a Josephson junction in series with the contact. Therefore, in contrast to recent STM data[10], a single gap of either s or $d_{x^2-y^2}$ symmetry cannot explain the data presented here.

Multigap Features

We now turn to examine spectra taken from a different region of the same polycrystalline sample, that do not show the zbcp such as the example shown in figure 3. This data is unusual because it shows two clear peaks in the conductance curves at $V \approx 4.5$ mV and $V \approx 8.6$mV suggesting the possible observation of multigap superconductivity. Note that at 4.2 K, only one broad gap feature is evident in the point contact spectrum at ~5.4-6meV, although there is a weak shoulder on the data at $V \sim 8.8$meV. As the temperature is increased to 5 K two peaks become clearly discernable in the spectra. (Note that the observation of a change in the spectra at low temperature is a common feature in this material and might be associated with magnetic ordering of Tb at low temperatures. This effect is not discussed further here). If both peaks are associated with superconducting gaps, then, for a sample where $T_c = 50$K, the BCS ratio ($2\Delta/kT_c$) is 2.1 and 3.9 respectively. In figure 4 we fit the 5K data shown in figure 3 using the 2 s-wave order parameter analysis[17]. This gives $\Delta_1 = 5.0$ meV and $\Delta_2 = 8.8$ meV with $Z = 0.37$, broadening parameter $\Gamma = 0.65$ meV and a weighting factor between the two gaps of $f = 0.42$. Figure 4 also shows typical example fits to other spectra where unlike curve (a) two gaps are not well resolved but the spectra peak is broad, including (c) the 4.2K spectrum in figure 1a. It is interesting to note that for similar temperatures, all spectra give similar energy gap values and similar values for the breadth of this peak feature, whether the peaks are resolved or not.

Dimensionality

Up until this point we have presented data that addresses the symmetry and multigap nature of these materials. In order to examine the dimensionality of the superconductivity we extend the experiments to include interrogation using a Nb tip as a function of temperature above and below $T_c$(Nb). The results shown in figure 5 are preliminary but interesting. The theory to describe the conductance resulting from a point contact between two superconducting materials was developed by Octavio et al[26] and results in the observation of sub-gap structure which appears as peaks in the conductance of the junction[27]. Previously this method was used for the two gap superconductor, $MgB_2$, to demonstrate that, although in principle sub-gap features should appear at $\Delta_i/n$ (n=1,2,3 etc and i = $\Delta_{Nb}$, $\Delta_{\sigma,\pi}$) and $(\Delta_{Nb}+\Delta_{\pi,\sigma})/m$ (where m= 1,3,5 etc), only the gap $\Delta_\pi$ associated with the three dimensional Fermi surface participated in the process[28].
Measuring the point contact spectrum of the $TbFeAsO_{0.9}F_{0.1}$ with a Nb tip at T = 11K (normal state of Nb) and T < $T_c$(Nb) (9.2K) ought to give some indication of features associated with the superconducting energy gaps that participate in the development of sub-gap structure. Figure 5 shows the conductance spectra of the $TbFeAsO_{0.9}F_{0.1}$ with a Nb tip at 5.6K and 10K. Figure 5(a) shows that in addition to a prominent zbcp associated with the Josephson current across the Nb- oxypnictide junction, significant enhancement in the conductance is also observed in two bias regions at V ≤ |8mV| and |8mV| < V ≤ |30mV|. At low bias voltages there are clear sub-gap structure peaks which move to increasingly lower bias as the temperature is raised, as shown in figure 5(b). By tracking these peaks in the conductance as a function of temperature the first peak is associated with $\Delta_{Nb}$ (1.5meV). There are clearly several additional peaks at higher voltages (a broad feature at 2.0 ± 0.4 mV, 3.5±0.2mV, 4.6±0.4mV) that indicates that coupling between the two superconductors exists. By assuming gap values of $\Delta_1$ = 5.0meV and $\Delta_2$ = 8.8 meV, we can tentatively assign these peaks to $\Delta_1/3$, $(\Delta_2+\Delta_{Nb})/5$, or $(\Delta_1+\Delta_{Nb})/3$; $(\Delta_2+\Delta_{Nb})/3$ and $\Delta_1$ or $\Delta_2/2$ respectively. Clearly these results are promising but preliminary.

**4. Discussion**
Since the literature on the oxypnictides is extensive we summarise the position with respect to PCAR data to date. There have been attempts to explain the experimental PCAR results based on an unconventional order parameter either of the s± type proposed originally by Mazin et al[4], or the $s_{xy}$ type proposed by Seo et al.[29, 30, 31, 32]. For completion we include a summary of the PCAR and STM results on a variety of materials systems in table 1. One key observation that led some early PCAR and STM studies to conclude that the order parameter was d-wave was the presence in the conductance spectra of a zero bias conductance peak (zbcp) or V shaped zero bias feature[10,12,13,14,33], although other studies showed spectra that could be fitted well with a simple, s-wave model[9], or which had no zero bias anomaly[8,33]. Indeed, it is perfectly feasible that zbcp can appear in spectra from polycrystalline superconductors without d-wave symmetry by Josephson effects occurring in series with the contact[24]. The presence of a zbcp is therefore not necessarily indicative of a d-wave order parameter. The need for caution in interpreting this zbcp as d-wave has been further emphasised by the theoretical work investigating the consequences to tunnelling measurements of the s± (sπ) state. In these works it was not only shown that the zbcp can be a natural consequence of the s± (sπ) state[30], (although there has recently been discussion about this point[34]), it was also shown that the V shaped conductance can arise from increased impurity scattering[31] or electron doping[32]. Even if Josephson effects can be ruled out as a cause of a zbcp in a spectrum as we have done here, the presence of a zbcp is consistent with both d-wave and s± (sπ) order parameters.

**Table 1 Summary of PCAR and STM results on the gap structure of a variety of oxyarsenide samples. * indicates the current study.**

| Composition | Technique | Polycrystal/ Single crystal | $T_c$ (K) | Number of gaps observed | v-shape or zbcp | $2\Delta/kT_c$ of gaps observed |
|---|---|---|---|---|---|---|
| $(Ba_{0.55}K_{0.45})Fe_2As_2$ | PCAR [15] | single | 23-27 | 2 | No | 1.8-4.6 8.3-10.2 |
| $Sr_{1-x}K_xFe_2As_2$ | STM [12] | Poly | 32 | 1 | V | 7.25 |
| $SmFeAsO_{0.85}$ | STM [10] | Poly | 52 | 1 | V shape | 3.55-3.8 |
| $SmFeAsO_{0.85}F_{0.15}$ | PCAR [9] | poly | 42 | 1 | No | 3.68 |
| $SmFeAsO_{0.9}F_{0.1}$ | PCAR [14] | poly | 51.5 | 2 | Zbcp | 1.7 4.5 |
| $LaFeAsO_{0.9}F_{0.1}$ | PCAR [13] | Poly | 28 | 1 | Zbcp | 3.35 |
| $LaFeAsO_{0.9}F_{0.1}$ | PCAR [8] | Poly | 27 | 2 | No | 2.4-3.95 |

| | | | | | | 8.4-10.3 |
|---|---|---|---|---|---|---|
| NdFeAsO$_{0.9}$F$_{0.1}$ | PCAR [11] | poly | 51 | 1 | No | 1.36-3.2 |
| NdFeAsO$_{0.85}$ | PCAR [33] | poly | 45 | 1 | Zbcp | 3.57 |
| TbFeAsO$_{0.9}$F$_{0.1}$ | PCAR * | Poly | 50 | 2 | sometimes zbc | 2.1 |
| | | | | | | 3.9 |

Although the presence of a zbcp does not explicitly reveal the symmetry of the order parameter(s) PCAR can nonetheless reveal important information concerning the number of gaps in the material, provided these gaps are resolvable compared to kT. Theoretically it has been predicted within the s± (sπ) model that the magnitude of the gaps on the hole Fermi surface and the electron Fermi surface are of similar size. However, the gaps on each Fermi surface are also predicted to be doping dependent such that $\Delta_e N_h \approx \Delta_h N_e$[30,31,35]. It is therefore not clear whether there ought to be two measurably distinct gaps or not. Nevertheless as can be seen in table 1 several groups have attempted to fit PCAR spectra to multiple gap models. The observations we have made are consistent with those of Wang et al, who observed two features at 2Δ/kT$_c$ of ~1.7 and ~4.5[14] in SmFeAsO$_{0.9}$F$_{0.1}$. They are also within the range of the lower gap feature observed by Daghero et al[36].

Although there may be differences between the hole doped 122 systems and the electron doped 1111 systems, Szabo et al.,[15] and Gonnelli et al.,[8] have observed additional features, usually as clearly resolved high bias shoulders on the spectra which result in a gap like feature at about 2Δ/kT$_c$ = 7.5 – 9 in 122 and 1111 compounds, respectively. We see shoulder like features in our data that are at similar energy scales but we have not attempted to fit them as they are generally broad and poorly resolved. It is interesting to note that it is this upper shoulder that is consistent with superconducting energy gap values found by ARPES[6] and FTIR [37].

In conclusion, we have summarised the observations from the literature using the PCAR method and compared these observations to our own data. There appears to be some convergence in understanding and some aspects of the data are consistent between groups using the PCAR method. Clearly much more work needs to be done in order to understand the role of doping and impurities in these materials before we will fully understand gap symmetry, gap number and gap dimensionality. Nevertheless in the present contribution we appear to have ruled out the zbcp in our PCAR data on TbFeAsO$_{0.9}$F$_{0.1}$ as being directly related to a Josephson junction in series with the tip junction or to a model invoking a single d-wave order parameter. Most groups now favour a multigap scenario and several groups are fitting the high bias feature and interpreting this as a superconducting energy gap. In our case we have seen clear evidence for two well resolved spectroscopic features which are usually encompassed by one broader feature. These features may result from sample inhomogeneity and/or orientational dependence due to the polycrystalline nature of the material. Preliminary results from the superconducting Nb tip spectroscopy data suggest that both gaps individually participate in the sub-gap structure, although further work on single crystals would be needed to confirm this statement.

Arxiv: 0812.0977; New J Phys, **11**, 025015 (2009)

**Figure captions**:

Figure 1: (a) Temperature dependence of a contact between TbFeAsO$_{0.9}$F$_{0.1}$ and an Au tip showing a zbcp. The spectra have been normalised and offset for clarity. (b) Fit to the 4.2K data assuming a d$_{x2-y2}$ order parameter and $\Delta$ = 8.5meV, Z = 0.4, $\alpha$ = 0.21rad and $\Gamma$ = 0.97meV. (c) magnetic field dependence of a contact at 8K showing a zbcp. The spectra have been normalised and offset for clarity.

Figure 2: The ratio of the height of the zbcp conductance to the height of the peak at finite bias, as a function T (circles). Blue triangles are the result of increased thermal broadening to the 4.2K fit. The dashed line indicates the behaviour of the zbcp height if $\alpha$ changes from 0.21 to 0.19 or Z changes from 0.40 to 0.48

Figure 3: Temperature dependence of a contact between TbFeAsO$_{0.9}$F$_{0.1}$ and an Au tip showing no zbcp but multiple gap like features, data have been normalised and offset for clarity.

Figure 4: Fits to three contacts from different regions of the sample using a 2 s-wave model. From top, (a) 5.6K $\Delta_1$ = 5.0meV and $\Delta_2$ = 8.8meV, Z = 0.37, $\Gamma$ = 0.65 meV; (b) 10K $\Delta_1$ = 4.6meV and $\Delta_2$ = 8.6meV, Z = 0.44, $\Gamma$ = 2.58 meV; (c) 4.2K, $\Delta_1$ = 6.1meV and $\Delta_2$ = 8.0meV, Z = 0.49, $\omega$ = 2.56 meV. Spectra have been normalised and offset for clarity. Data points are indicated by symbols, fit is shown by the thin red line.

Figure 5: (a) Conductance curve for a TbFeAsO$_{0.9}$F$_{0.1}$ - Nb tip contact at 5.6K (blue up- triangle) and 10.7K (red circle). (b) Shows a magnified low bias region for the same contact at 5.0 K (navy square), 5.6K (blue up-triangle), 5.9 K (cyan down-triangle), 10.7 K (red circle). Spectra in (b) have been normalised and offset for clarity. The arrow indicates the direction of increasing temperature.

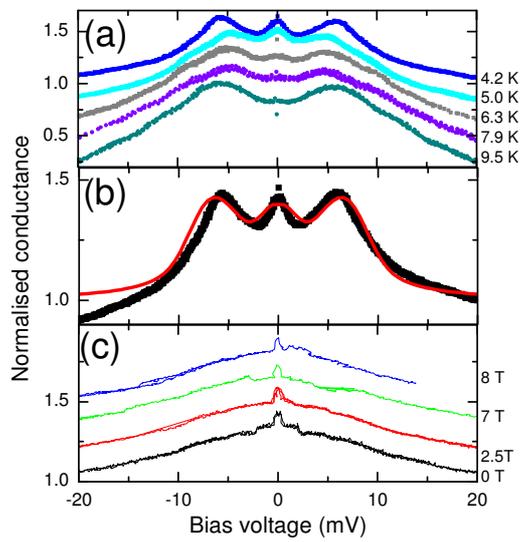

Figure 1

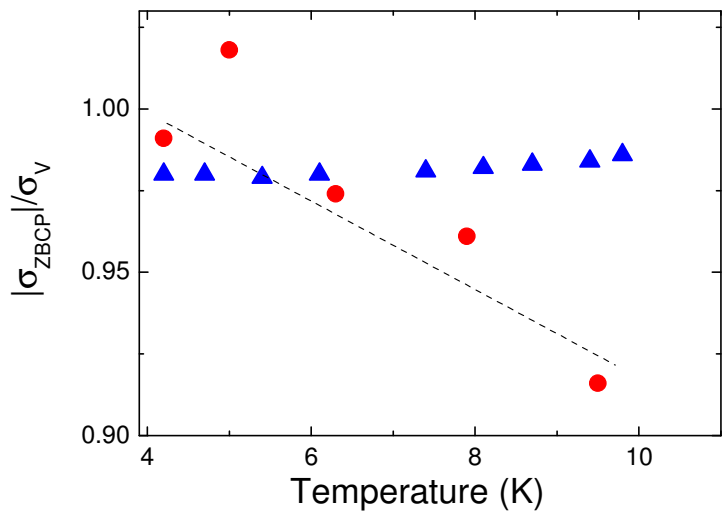

Figure 2

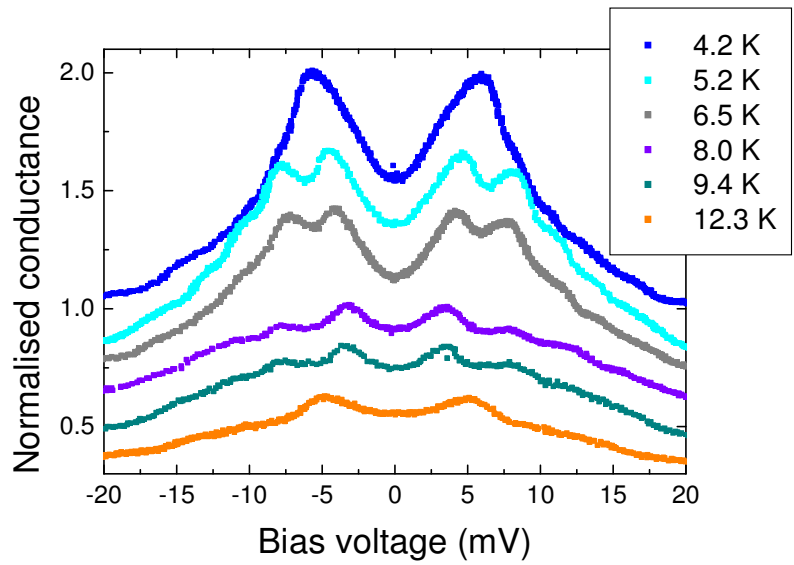

Figure 3

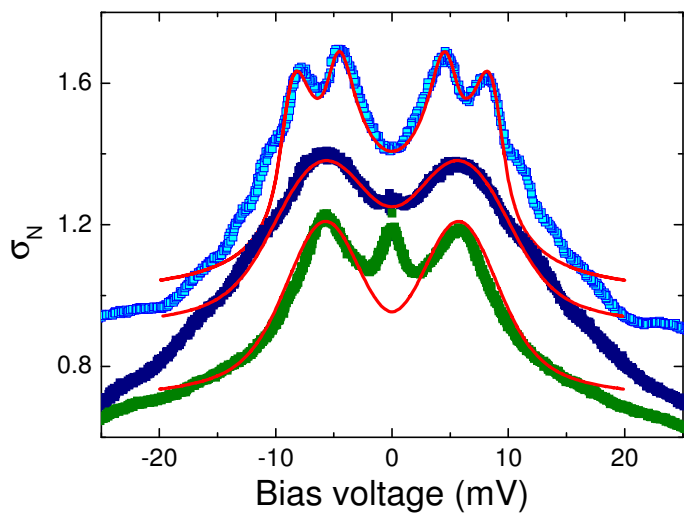

Figure 4

Arxiv: 0812.0977; New J Phys, **11**, 025015 (2009)

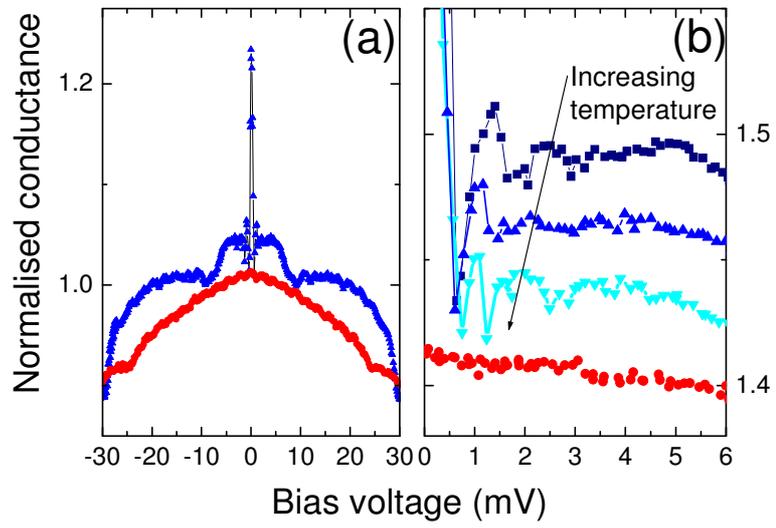

Figure 5